\documentclass[conference, twocolumn]{IEEEtran}

\usepackage[usenames,dvipsnames]{xcolor}
\usepackage{amsmath,amssymb,amsfonts,tikz,pgfplots,bm}
\usepackage{cite,booktabs,tabularx}
\usepackage{algorithmic}
\usepackage{graphicx}
\usepackage{textcomp}
\usepackage{verbatim}
\usepackage{amsmath}
\usepackage[ruled,vlined]{algorithm2e}
\usepackage{nicefrac}
\usepackage[inline]{enumitem}
\usepackage{array}
\usepackage{balance}
\usepackage{fixltx2e}
\usepackage{standalone}
\usetikzlibrary{external}

\newcommand{\len}{\ensuremath{N}}
\newcommand{\leni}{\ensuremath{n}}
\newcommand{\mem}{\ensuremath{m}}
\newcommand{\dimi}{\ensuremath{k}_{\text{CC}}}

\newcommand{\leno}{\ensuremath{N_\mathsf{o}}}
\newcommand{\lenob}{\ensuremath{N_\mathsf{o}^\mathsf{b}}}
\newcommand{\dimo}{\ensuremath{K}}
\newcommand{\numS}{\ensuremath{M}}

\newcommand{\rateo}{\ensuremath{R_\mathsf{o}}}
\newcommand{\ratei}{\ensuremath{R_\mathsf{i}}}

\newcommand{\D}{\mathsf{D}}
\newcommand{\I}{\mathsf{I}}

\newcommand{\Ins}{{\tt{{Ins}}}}
\newcommand{\Del}{{\tt{{Del}}}}
\newcommand{\Sub}{{\tt{{Sub}}}}
\newcommand{\Trans}{{\tt{NoErr}}}

\newcommand{\calD}{\mathcal{D}}

\let\vv\v

\newcommand{\e}{\ensuremath{\mathrm{e}}}

\newcommand{\s}{\ensuremath{\boldsymbol{s}}}
\renewcommand{\u}{\ensuremath{\boldsymbol{u}}}
\newcommand{\w}{\ensuremath{\boldsymbol{w}}}
\newcommand{\bx}{\ensuremath{\boldsymbol{x}}}
\newcommand{\bz}{\ensuremath{\boldsymbol{z}}}

\newcommand{\x}{\ensuremath{\boldsymbol{x}}}

\newcommand{\y}{\ensuremath{\boldsymbol{y}}}

\renewcommand{\v}{\ensuremath{\boldsymbol{v}}}

\newcommand{\Al}{\ensuremath{\Sigma}}

\definecolor{ColorIGMDDSR}{rgb}{0,0.49804,0}

\def\bwidth{3.9em}
		
\usetikzlibrary{positioning,calc}
\tikzstyle{block} = [rectangle, draw, rounded corners, text centered, minimum height=1.5em,fill=blue!7!]

\tikzstyle{seqblock} = [rectangle, draw, text centered, minimum height=1.6em,fill=blue!7!, minimum width=\bwidth, outer sep=0]

\def\approxprop{%
	\def\p{%
		\setbox0=\vbox{\hbox{$\propto$}}%
		\ht0=0.6ex \box0 }%
	\def\s{%
		\vbox{\hbox{$\sim$}}%
	}%
	\mathrel{\raisebox{0.7ex}{%
			\mbox{$\underset{\s}{\p}$}%
	}}%
}

\tikzstyle{round} = [circle, draw, text centered, minimum height=1.5em, fill=blue!7!]

\def\approxprop{%
	\def\p{%
		\setbox0=\vbox{\hbox{$\propto$}}%
		\ht0=0.6ex \box0 }%
	\def\s{%
		\vbox{\hbox{$\sim$}}%
	}%
	\mathrel{\raisebox{0.7ex}{%
			\mbox{$\underset{\s}{\p}$}%
	}}%
}

\usepackage{soul}

\newcommand{\field}[1]{\ensuremath{\mathbb{F}_{#1}}}

\newcommand{\outq}{\ensuremath{q_\mathsf{o}}}

\newcommand{\ocw}{\ensuremath{\boldsymbol{w}}}

\IEEEoverridecommandlockouts

\begin{document}

\title{Achievable Information Rates and Concatenated Codes for the DNA Nanopore  Sequencing Channel}

\author{\IEEEauthorblockN{{\bf Issam Maarouf}\IEEEauthorrefmark{1}, {\bf Eirik Rosnes}\IEEEauthorrefmark{1}, and {\bf Alexandre Graell i Amat}\IEEEauthorrefmark{2}}

\IEEEauthorblockA{
	\IEEEauthorrefmark{1}Simula UiB, N-5006  Bergen,  Norway
}

\IEEEauthorblockA{
	\IEEEauthorrefmark{2}Department of Electrical Engineering, Chalmers University of Technology, SE-41296 Gothenburg, Sweden
}

\thanks{The work of A. Graell i Amat was supported by the Swedish Research Council under grant 2020-03687.}} %

\maketitle

\begin{abstract}
The errors occurring in DNA-based storage  are correlated in nature, which is a direct consequence of the synthesis and sequencing processes. %
In this paper, we consider the memory-$k$ nanopore channel model recently introduced by Hamoum \emph{et al.}, which models the inherent memory of the channel. We derive the maximum a posteriori (MAP) decoder for this channel model. %
The derived MAP decoder allows us to compute achievable information rates for the true DNA storage channel assuming a mismatched decoder matched to the memory-$k$ nanopore channel model, and quantify the loss in performance assuming a small memory length\textemdash and hence limited decoding complexity. Furthermore, the derived MAP decoder can be used to design error-correcting codes tailored to the DNA storage channel. We show that a  concatenated coding scheme with an outer low-density parity-check code and an inner convolutional code %
yields excellent performance.
\end{abstract}

\IEEEpeerreviewmaketitle

\section{Introduction} \label{sec:introduction}

Storing data in deoxyribonucleic acid (DNA) promises unprecedented density and durability and is seen as the new frontier of data storage. Recent experiments have already demonstrated the viability of DNA-based data storage \cite{church_next-generation_2012, grass_robust_2015,yazdi_portable_2017}. 

The DNA storage channel suffers from multiple impairments and constraints due to the synthesis and sequencing of DNA and the limitations of current technologies. In particular, errors in the form of insertions, deletions, and substitutions (IDS) occur. This has spurred a great deal of research on devising coding schemes for the DNA storage channel. 

While the literature on error-correcting coding for the DNA storage channel is abundant, most works have considered
a simplified and unrealistic channel model with a small and fixed number of insertions and/or deletions, i.e., an \emph{adversarial} channel. Fewer works have considered a more realistic probabilistic channel model in which errors occur with a given probability, e.g.,  \cite{Maa22,HeckelShomorony_Foundations_2022,Srinivasavaradhan2021TrellisBMA,davey_reliable_2001,briffa_improved_2010,inoue_adaptive_2012,shibata_design_2019,Koremura2020,SakogawaMAP_IDS_2020,Shibata2020,Pfister2021,mansour_convolutional_2010,buttigieg_improved_2015}. 
These works usually assume independent and identically distributed (i.i.d.) errors. In DNA storage, however, IDS errors are not i.i.d., but the channel has memory %
\cite{Hamoum_IDSnanopore2021,McBain2022FiniteStateSC}. Potentially, the memory is as large as the whole DNA sequence.

In \cite{Hamoum_IDSnanopore2021}, the authors proposed a statistical model of the DNA storage channel with nanopore sequencing %
based on a Markov chain that models the inherent memory of the channel. In particular, it builds on the fact that in the MinION technology strands traverse a nanopore
nucleotide by nucleotide, and an electrical signal is generated  for every group of $k \geq 1$ successive nucleotides, called $k$mers. Let $\bx$ be the DNA strand to be synthesized and $\bz$ be the sequence of channel events, where  $z_i\in\{\tt{insertion}, \tt{deletion}, \tt{substitution}, \tt{no}\,\tt{error}\}$. The key idea in \cite{Hamoum_IDSnanopore2021} is then to assume that $z_i$ depends on the symbols $x_{i-k+1},\ldots,x_i$ and the previous event $z_{i-1}$ and 
estimate the probabilities $p(z_i|x_{i-k+1},\ldots,x_i,z_{i-1})$ from experimental data. We refer to this channel model as the memory-$k$ nanopore channel model.

The starting point of this paper is the work \cite{Hamoum_IDSnanopore2021}. Our main contributions are as follows.
\begin{itemize}
\item We derive the  optimum maximum a posteriori (MAP) decoder for the memory-$k$ nanopore channel model. %
The complexity of the decoder increases exponentially with $k$. Based on this decoder, we derive  achievable information rates (AIRs).
\item The AIR for the memory-$k$ nanopore channel  can be seen as an AIR for the true DNA storage channel of a \emph{mismatched} decoder (i.e., a decoder that is not matched to the true channel) that assumes that the channel is a memory-$k$ nanopore channel. We show that for increasing $k$, the AIR improves\textemdash meaning that the decoder is better matched to the true channel\textemdash and eventually saturates. This allows us  to quantify the trade-off between decoding complexity and performance loss incurred by the suboptimal decoder.
\item The derived MAP decoder can be used to design error-correcting coding schemes tailored to the memory-$k$ nanopore sequencing channel. In particular, we consider the concatenated coding scheme proposed in \cite{Maa22} and multiple reads of the DNA strand, and we optimize the inner and outer code. %
We show that the concatenated coding scheme of \cite{Maa22} yields excellent performance at rates close to the AIRs. 

\item We validate the AIRs and simulation results of the memory-$k$ nanopore channel model   using  the dataset of DNA reads in \cite{Srinivasavaradhan2021TrellisBMA} obtained using Oxford Nanopore Technologies (ONT) sequencing with the MinION technology.

\end{itemize}

\section{The Nanopore Sequencing Channel Model}
\label{sec:system}
\begin{figure}[t]
	\centering

    \includegraphics[scale = 0.9]{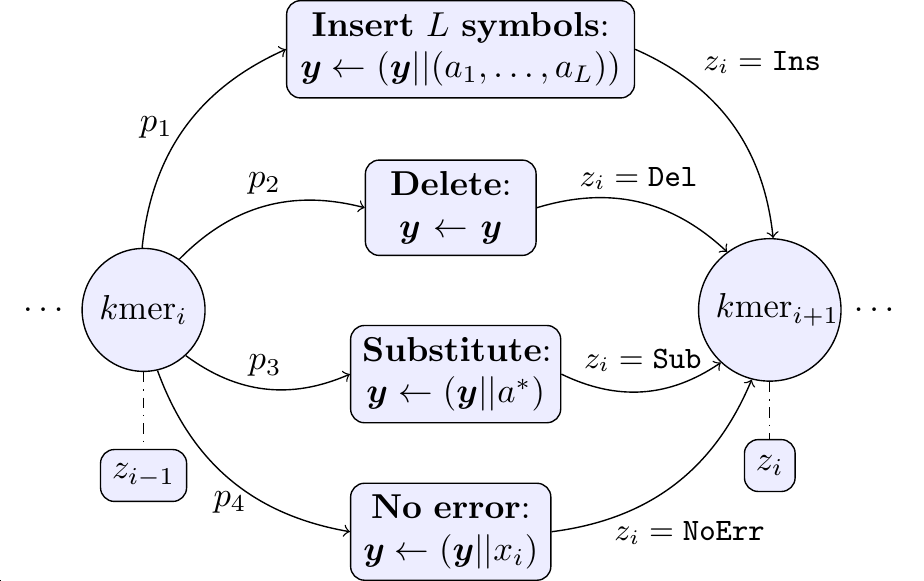}
	\vspace{-1ex}
	\caption{The memory-$k$ nanopore   channel in \cite{Hamoum_IDSnanopore2021}.}
	\vspace{-2ex}
	\label{fig:ids:channel}
\end{figure}

We consider the memory-$k$ nanopore channel model introduced in \cite{Hamoum_IDSnanopore2021}. The key property of this channel model is that it removes the assumption of i.i.d. IDS errors. In particular, the occurrence of an error event, or a correct transmission, depends on the previous event (an insertion, deletion, substitution, or no error event), and  previous channel input symbols. Let  $\x = (x_1,\ldots ,x_\len)$, $x_i \in \Al _4 = \{0,1,2,3\}$, be the  channel input sequence and $\y = (y_1,\ldots,y_{\len'})$ the channel output sequence, and  define $x_{N+1}=0$. Note that $\len'$ is random and depends on the number of insertion and deletion events, i.e., $N'\neq N$ in general. Also, let $k\text{mer}_i = (x_{i -k + 1},\dots,x_i)$ be the $k$mer at time instant $i \geq k$, while for $i<k$, since no complete $k$mer is available, we use $k\text{mer}_i=x_i$.  %
The channel model is  depicted in Fig.~\ref{fig:ids:channel}. Here, we denote the event associated with $k\text{mer}_{i+1}$ by $z_i$, where $z_i$ represents an insertion, deletion, substitution, or no error  acting on symbol $x_i$ when it is to be transmitted. Hence, $z_i \in \{\Ins, \Del, \Sub, \Trans \}$. The transition probabilities $p_1$, $p_2$, $p_3$, and $p_4$ in the figure are of the form $p(z_i|k\text{mer}_i, z_{i-1})$, and their value depends on the $k$mer. In the event of a deletion, the symbol $x_i$ in $k\text{mer}_i$ will be deleted and nothing will be appended to $\y$, while when a no error event occurs, $\y$ will be appended with $x_i$. Furthermore, in the event of an insertion, i.e., $z_i = \Ins$, the channel will insert $L > 1$ symbols $a_1,\dots,a_{L}$, where the length $L$ is random. Each symbol $a_j$ will be uniformly picked from $\Al_4$. In this case, the channel appends $\y$ with $(a_1,\dots,a_{L})$. This also applies to a substitution event where the symbol $a^*$ is uniformly picked from $\Sigma_4 \setminus \{x_i\}$ and appended to $\y$ instead of $x_i$. Moreover, we define $p(L| k\text{mer}_i, z_i = \Ins)$ as the probability of inserting $L$ symbols given $k\text{mer}_i$ and an insertion event, and $p(a^*| k\text{mer}_i, z_i = \Sub)$ as the probability of $x_i$ being substituted by $a^*$ given $k\text{mer}_i$ and a substitution event. In all these scenarios, the channel always moves from state $k\text{mer}_i$ to state $k\text{mer}_{i+1}$. After the last symbol $x_{\len}$ has been processed by the channel, the channel outputs $\y$.
%

%

%
\subsection{Channel Transition Probability Estimation} \label{sec:model_build}
The transition probabilities of the channel model can be estimated using a DNA storage dataset $\calD$ containing  input (before synthesis) and the corresponding multiple output (after sequencing) sequences. 

Let $\x^{(l)}$ denote the $l$-th  input sequence and $\y^{(l)}_1, \ldots, \y^{(l)}_{\numS_l}$ the corresponding $\numS_l \geq 1$ output sequences from the dataset $\calD$.
The method we use to estimate the transition probabilities is as follows. 
For each pair $(\x^{(l)}, \y_j^{(l)})$, we compute the edit distance between the two vectors using a lattice (see \cite[Sec.~III]{Maa22},\cite{bahl_decoding_1975}).  By backtracking from the end of the lattice, a corresponding  sequence of events $\bz_j^{(l)}$ can be identified. In case of ties, one event is chosen uniformly at random.     
Finally, we estimate the probabilities $p(z_i=z|k\text{mer}_i=k\text{mer}, z_{i-1}=z')$ by a simple counting argument. 
In particular, let %
\begin{align*}
\mathcal{S}_{z,z',k\text{mer}}^{(l,j)}=\big\{k \leq u < N:\, & (x^{(l)}_{u-k+1},\ldots,x^{(l)}_{u})=k\text{mer}, \\
&  (z_{j,u-1}^{(l)},z_{j,u}^{(l)})=(z',z) \big\}\,.
\end{align*} 
Then, for $k \leq i < N$, $p(z_i=z|k\text{mer}_i=k\text{mer}, z_{i-1}=z')$ is estimated by 
\begin{align*}
\frac{\sum_{l=1}^{|\calD|}\sum_{j=1}^{\numS_l} |\mathcal{S}_{z,z',k\text{mer}}^{(l,j)}|}{\sum_{z \in \{\Ins,\Del,\Sub,\Trans \}} \sum_{l=1}^{|\calD|}\sum_{j=1}^{\numS_l} |\mathcal{S}_{z,z',k\text{mer}}^{(l,j)}|}\,.
\end{align*}
As a consequence, the channel parameters do not vary with $i$ for $k \leq i < N$.  For $1 < i < k$, we estimate the channel transition probabilities in a similar manner as above, but considering that $k\text{mer}_i=x_i$. Moreover, as in  \cite{Hamoum_IDSnanopore2021}, we consider $i=1$ (the beginning) and $i=N$ (the end) separately from the middle ($1 < i < N$) when estimating the channel parameters as  the probability to get an error is typically higher at the beginning and at the end compared to the middle.

\begin{figure}[t]
	\centering

   \includegraphics[scale = 0.75]{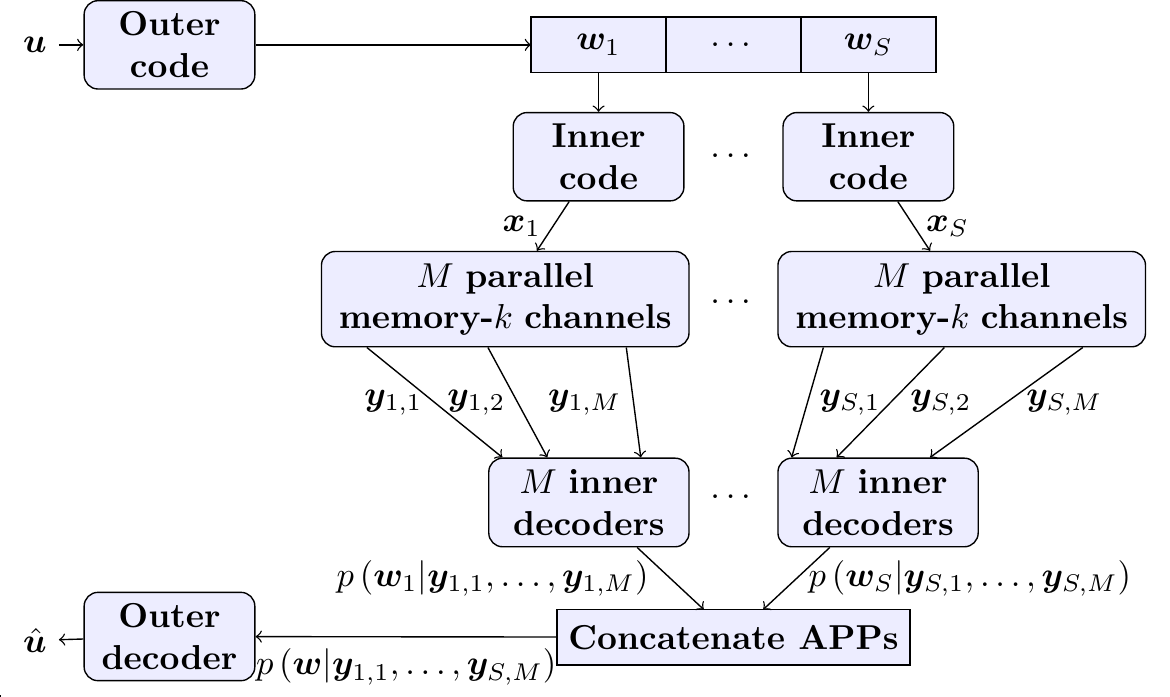}
  \vspace{-3ex}
	\caption{System model including the coding scheme and the memory-$k$ nanopore channel model depicted in Fig.~\ref{fig:ids:channel}.} 
	\label{fig:system:model}
	\vspace{-2.0ex}
\end{figure}

\subsection{Multiple Reads}

We consider the relevant scenario in which the sequencing process outputs multiple reads ($M$) of the DNA strand. In particular, we model this by assuming that a strand is transmitted over $M$ independent channels.

\section{Coding Scheme and System Model}
We consider the concatenated coding scheme and the low-complexity \emph{separate decoding} scheme proposed in  \cite{Maa22}, which decodes separately each noisy strand and  combines a posteriori probabilities (APPs) from all reads before passing them to the outer decoder. The primary goal of the inner code is to maintain synchronization with the transmitted sequence and provide likelihoods to the outer decoder. The outer code then corrects remaining errors.

The system model is shown in Fig.~\ref{fig:system:model}. We consider a low-density parity-check (LDPC) code for the outer code and a convolutional code for the inner  code. The information data $\u = (u_1, \ldots, u_\dimo)$, $u_i \in \field{\outq}$, of length $\dimo$, is first encoded by an $[\leno,\dimo]_{\outq}$ outer LDPC code over $\field{\outq}$ into the codeword $\ocw = (w_1,\dots,w_{\leno})$, $w_i \in \field{{\outq}}$, where  $\field{{\outq}}$ is a binary  extension field of size $2^{\dimi}$ and $\dimi$ is the (binary) dimension of the inner convolutional code. 
 Since current sequencing technologies cannot handle long sequences, the codeword $\ocw$ is split into $S$  subsequences $\ocw_1,\dots, \ocw_S$, each of length $\lenob = \nicefrac{\leno}{S}$, where $\ocw_j = (w_{(j-1)\lenob + 1},\dots, w_{j\lenob})$. Each $\ocw_j$ is encoded, with an optional offset \cite{Maa22}, by an $(\leni,\dimi, \mem)_4$ inner convolutional code of binary dimension $\dimi$, memory $m$, and output alphabet $\Al _4$ into the inner codeword $\bx_j$ of length $N=(\lenob + \mem)n$, where $\mem$ is due to terminating the convolutional code to the all-zero state. %
The overall code rate is given as $R = \rateo \ratei = \nicefrac{\dimo \dimi}{S\len}$, where  $\rateo = \nicefrac{\dimo}{\leno}$ is the rate of the outer code  and $\ratei = \nicefrac{\dimi \lenob}{\len}$ the rate of the inner code.
 
Each $\x_j$ is transmitted independently over  $\numS$ identical memory-$k$ nanopore channels in order to model the multiple copies of a DNA strand at the output of the sequencing process. %
At the receiver, each noisy read  $\y_{j,1},\ldots,\y_{j,\numS}$ for  transmitted sequence $\x_j$ is decoded separately with a MAP inner decoder. The APPs at their output are then combined to provide approximate APPs for $\w_j$. The (approximate) APPs for the symbols in $\w=(\w_1,\ldots,\w_S)$ are concatenated and then passed to the outer decoder. The outer decoder uses these APPs to decide on an estimate $\hat{\u}$ of $\u$.
\section{Symbolwise MAP Decoding for Channel/Inner Code (Inner Decoding)} \label{sec:innerdec-one}

In this section, we derive the optimum (MAP) decoder for the combination of the memory-$k$ nanopore channel model and the inner code. To this end, we use the fact that the combination of the channel and the inner code can be seen as a hidden Markov model (HMM) by introducing a \emph{drift} variable as in \cite{davey_reliable_2001}. The drift $d_i$, $0 \leq i < \lenob + m$, is defined as the number of insertions minus the number of deletions that occurred before symbol $x_{ni+1}$ is to be acted on by the event $z_{ni + 1}$, while $d_{\lenob+m}$ is defined as the number of insertions minus deletions that occurred after the last symbol $x_{N}$ has been processed by the channel.  
Thus, by definition, $d_0 = 0$ and $d_{\lenob+m} = N'-N$, both known to the decoder. Adding the drift to the joint state of the channel and the inner code, i.e., to $(k\text{mer}_{i+1},z_i, s_i)$, where $s_i$ denotes the state variables of the inner convolutional code, gives the state variable $\sigma_{i} = (k\text{mer}_{i+1},z_i, s_i, d_i)$ of the HMM. 
In this HMM, a transition from time $i-1$ to time $i$ corresponds to a transmission of symbols $\x_{(i-1)n+1}^{in}$, where $\x_a^b = (x_a,x_{a+1},\dots, x_b)$. Further, when transitioning from a state with drift $d_{i-1}$ to a state with drift $d_{i}$, the HMM emits $n+d_{i}-d_{i-1}$ output symbols depending on both the previous and new drift.

In the following, to simplify notation, the subsequence index of  $\ocw$ and $\y$ is omitted and $w_i$ simply refers to the $i$-th symbol of an arbitrary subsequence $\ocw_j$, while $\y=(\y_1,\ldots,\y_{\numS})$ refers to the corresponding received sequences.

\subsection{Decoding for a Single Received Sequence} \label{sec:single_seq}

The APP for outer code symbol $w_i$ can be computed as 
 $p(w_i|\y) = \nicefrac{p(\y,w_i)}{p(\y)}$,
where the joint probability $p(\y,w_i)$ can be computed by marginalizing the trellis states, corresponding to the HMM, of the channel/inner code that correspond to symbol $w_i$. %
Then, we can write
	$p(\y,w_i) = \sum_{(\sigma,\sigma'):w_i} p(\y,\sigma,\sigma')$,
where $\sigma$ and $\sigma'$ are realizations of the random variables $\sigma_{i-1}$ and $\sigma_i$, respectively. Here, the summation is over all  pairs of states that correspond to symbol $w_i$. We can use the Markov property to decompose the probability $p(\y,\sigma,\sigma')$ into three parts as
\begin{align*}
	&p(\y,\!\sigma\!,\!\sigma') \!=\!
	p\!\left(\!\y_{1}^{(i\!-\!1)n+d}, \sigma\!\right)\!p\!\left(\!\y_{(i\!-\!1)n\!+d+\!1}^{in+d'}, \sigma'\big|\sigma\!\right)\!p\!\left(\!\y_{in\!+\!d'\!+1}^{\len'}\Big| \sigma'\!\right)\!.
\end{align*}
We abbreviate the first, second, and third term of the above equation with $\alpha_{i-1}(\sigma)$, $\gamma_i(\sigma,\sigma')$, and $\beta_i(\sigma')$, respectively. Then, the first and third term can computed recursively  as
\begin{align*}
	\alpha_i(\sigma') &= \sum_{\sigma}\alpha_{i-1}(\sigma) \gamma_{i}(\sigma,\sigma'), %
	\beta_{i-1}(\sigma) = \sum_{\sigma'}\beta_i(\sigma') \gamma_i(\sigma,\sigma'). \notag
\end{align*}
The  term $\gamma_i(\sigma,\sigma')$ (the branch metric) can be decomposed as 
\begin{align*}
\gamma_i(\sigma,\sigma') &= p(w_i) p(z'| k\text{mer}, z)\notag \\
&\quad\cdot p\left(\y_{(i-1)n+d+1}^{in+d'}, d'\big|d,s,s',z',z,k\text{mer}',k\text{mer}\right)\,, \label{eq:gamma}
\end{align*}
where $p(w_i)$ is the a priori probability of symbol $w_i$. For simplicity,   define the state variable $\zeta_i = (k\text{mer}_{i+1}, z_i)$. To compute $p(\y_{(i-1)n+d+1}^{in+d'}, d'\big|d,s,s',\zeta',\zeta)$, we need to consider each possible event for $z'$. For simplicity, we limit our derivation to $n = 1$. The case for general $n$ follows in a straightforward manner.
\begin{enumerate}
    \item $z' = \Ins$. 
    Then,
    \begin{align*}
        &p\left(\y_{(i-1)+d+1}^{i+d'}, d'\big|d,s,s',\zeta',\zeta\right)\\&=p\left(\y_{(i-1)+d+1}^{i+d'}, L = d'-d\big|d,s,s',\zeta',\zeta\right)\\
        &\quad \cdot  p\left(L = d'-d\big|d,s,s',\zeta',\zeta\right)\\
        &= \left(\frac{1}{4}\right)^L \hspace{-1ex}\cdot p(L|k\text{mer},z' = \Ins)\,.
    \end{align*}
    
    \item $z' = \Del$. 
    Then, $d' = d-1$, which means that $p(\y_{(i-1)+d+1}^{i+d'}, d' \neq d - 1\big|d,s,s',\zeta',\zeta) = 0$ and
        $p(\y_{(i-1)+d+1}^{i+d'}, d'= d - 1\big|d,s,s',\zeta',\zeta) = 1$.

    \item $z' = \Sub$. 
    Then, $d' = d$, which means that $p(\y_{(i-1)+d+1}^{i+d'}, d' \neq d\big|d,s,s',\zeta',\zeta) = 0$, and
    \begin{align*}
        &p\left(\y_{(i-1)+d+1}^{i+d'}, d'= d\big|d,s,s',\zeta',\zeta\right)\\
        &= \begin{cases}
		p\left(\y_{i+d}^{i+d} \big| k\text{mer},s,s',z'=\Sub\right) & \text{if } \y_{i + d}^{i+d} \neq \x_{i + d}^{i+d}\\
		0& \text{otherwise}\,.
		\end{cases}
    \end{align*}
    
    \item $z' = \Trans$. 
    Then, $d' = d$, which means that $p(\y_{(i-1)+d+1}^{i+d'}, d' \neq d\big|d,s,s',\zeta',\zeta) = 0$, and
    \begin{align*}
        &p\left(\y_{(i-1)+d+1}^{i+d'}, d'= d\big|d,s,s',\zeta',\zeta\right)\\
        &= \begin{cases}
		1 &\y_{i + d}^{i+d} = \x_{i + d}^{i+d}\\
		0& \text{otherwise}\,.
		\end{cases}
    \end{align*}
\end{enumerate}

\subsection{Decoding for Multiple Received Sequences}
\label{eq:separate}

We consider the separate decoding strategy for multiple received sequences $\y_1,\dots,\y_{\numS}$ proposed in \cite{Maa22}.  Following \cite{Maa22}, the APP  $p(w_i|\y_1,\dots,\y_{\numS})$ can be approximated as %
\begin{align*}
    p(w_i|\y_1,\dots,\y_{\numS}) \approxprop \frac{\prod_{j=1}^{\numS} p(w_i|\y_j)}{p(w_i)^{\numS-1}}\,,
\end{align*}
where  $p(w_i|\y_j)$ is computed as outlined in Section~\ref{sec:single_seq}. 
This decoder, although suboptimal, is efficient and practical for our scenario, as its complexity  grows linearly with $\numS$ \cite{Maa22}.\footnote{Alternatively, one may decode all $M$ reads jointly using a single inner decoder \cite{Maa22}. However, the complexity of this decoder grows exponentially with the number of sequences $\numS$, and becomes infeasible for $\numS > 2$.}

\subsection{Decoding Complexity}
\label{sec:Complexity}
Since the overall decoding complexity is dominated by the combination of the inner code and the channel, we will disregard the complexity of the  outer decoder in the complexity analysis. In order to limit the inner decoding complexity, we limit the drift $d_i$ to a fixed interval $[d_{\text{min}},d_{\text{max}}]$ and the number of insertions per symbol $L$ to $L_{\text{max}}$; recall also that $L>1$.  For simplicity, we limit our derivation to $n = 1$. The complexity of the BCJR algorithm on the joint trellis of the inner code and the channel is directly proportional to the number of trellis edges at each trellis section, which is upper bounded by $2^{\nu+\dimi} 4^{k+1} \Delta (\delta+1)$, where $\Delta=d_{\text{max}}-d_{\text{min}}+1$ is the number of drift states, $\delta = L_{\text{max}} + 1$ is the number of possible drift transitions, and $\nu$ is the number of binary memory elements of the convolutional encoder.
Hence, the  complexity of decoding  a single block is $(\lenob+m) 2^{\nu +\dimi } 4^{k+1} \Delta  (\delta+1)$, and the  complexity of separate decoding (for all $S$ blocks) becomes $S (\lenob+m) 2^{\nu + \dimi} 4^{k+1} \Delta  (\delta+1) \numS$. Note that the complexity increases exponentially with the channel memory $k$. 
\section{Achievable Information Rates}\label{sec:air}

The MAP decoder for the memory-$k$ nanopore channel (including the inner code) derived in Section~\ref{sec:innerdec-one} allows us to compute  AIRs for this channel. In particular, we compute \emph{BCJR-once} rates \cite{Kavcic2003BinaryII,muller_capacity_2004,soriaga_determining_2007}, defined as the symbolwise mutual information between the input of the channel and the log-likelihood ratios (LLRs) produced by a symbolwise MAP (i.e., optimum) detector.  
For a given  inner code, the BCJR-once rate, denoted by $R_{\text{BCJR-once}}$, is a rate achievable by an  outer code that does not exploit possible correlations between the LLRs and when no iterations between the inner and outer decoder  are performed.

The BCJR-once rate under separate decoding can be estimated as
$$R_{\text{BCJR-once}} \approx \ratei \log \outq + \frac{\ratei}{\lenob + m} \sum_{i=1}^{\lenob + m} \log\frac{\e^{L^{\text{BCJR-sep}}_i(w_i)}}{\sum_{a \in \field{\outq}}\e^{L^{\text{BCJR-sep}}_i(a)}}$$
by sampling an input sequence $\w$ and corresponding output sequence $\y = (\y_1,\dots,\y_\numS)$ and computing the (mismatched) LLRs $L^{\text{BCJR-sep}}_i(a) = \sum_{j=1}^{\numS} \ln \frac{q(w_i=a|\y_j)}{q(w_i=0|\y_j)}$, $a \in \field{\outq}$, where $q(w_i|\y_j)$ is a (mismatched) inner decoding metric.

BCJR-once rates can also be computed for the true DNA storage channel using a dataset of DNA traces by averaging over  pairs  of input and output sequences. In this case, given an inner code and a decoder that assumes that the channel has memory $k$, the BCJR-once rate is an AIR for the DNA storage channel of a \emph{mismatched} decoder where the inner decoder is matched to the memory-$k$ nanopore channel model. %

For both cases, we assume  separate decoding and an inner MAP decoder matched to the memory-$k$ nanopore channel model, i.e., using  $q(w_i|\y_j) = p(w_i|\y_j)$ where $p(w_i|\y_j)$ is computed as described in Section~\ref{sec:single_seq}.   %

%
%
%
%
%
%
%

%
%
%
%
%
%
%
%
%
%
%
%
%
%

\begin{table}[t]
 \centering
	\caption{Optimized Protographs Found by DE for Different $\numS$} \label{tab:DE_protographs}
	\vspace{-2ex}
    	\begin{tabular}{ccc}
    	\toprule 
        \numS &  $R = \rateo \ratei$ &  Protograph  \\ %
         \midrule
         $1$ &  $\nicefrac{8}{10} \cdot \nicefrac{3}{2} = \nicefrac{6}{5}$ & $\left(\begin{smallmatrix} 1 & 2 & 0 & 1 & 2 & 2 & 1 & 1 & 1 & 3 \\ 2 & 0 & 3 & 2 & 1 & 0 & 1 & 2 & 2 & 0\end{smallmatrix}\right)$ \\[.2cm] 
         $2$ & $\nicefrac{7}{8} \cdot \nicefrac{3}{2} = \nicefrac{21}{16}$ & $\left(\begin{smallmatrix} 2 & 2 & 3 & 3 & 2 & 3 & 3 & 3\end{smallmatrix}\right)$\\[.2cm] 
         $5$& $\nicefrac{14}{15} \cdot \nicefrac{3}{2} = \nicefrac{7}{5}$ &$\left(\begin{smallmatrix} 3 & 2 & 3 & 3 & 2 & 2 & 3 & 3 & 2 & 2 &3 &3 &3 &2 &3\end{smallmatrix}\right)$ \\  \bottomrule
    	\end{tabular}
    \vspace{-3ex}
\end{table}

\section{Concatenated Coding Scheme Design} \label{sec:codes}

\subsection{Inner Code} \label{sec:inner_code}
For the inner code we use the $(1,1,2)_4$ convolutional code with generator polynomial  $g = [5,7]_{\text{OCT}}$ and punctured in order to have a higher rate. In particular, we use the puncturing matrix $\bm P = \left( \begin{smallmatrix} 1 & 0 & 1 \\ 1 & 1 & 0\end{smallmatrix}\right)$, which gives an inner code rate of $\ratei = \nicefrac{3}{2}$ (in bits per DNA symbol).  %
Moreover, we add a pseudo-random sequence to the output of the inner code.

\subsection{Outer Code}

We consider a protograph-based binary LDPC code as the outer code, which we optimize (also in terms of code rate) via density evolution (DE) using the algorithm proposed in \cite{Kavcic2003BinaryII}
for the memory-$5$ nanopore channel model for $M=1$, $2$, and $5$. %
Moreover, we set $\len = 110$ as the dataset in  \cite{Srinivasavaradhan2021TrellisBMA} contains input sequences with this length. The optimized protographs, when limiting the entries to at most $3$, are shown in Table~\ref{tab:DE_protographs}. 
%
%
%
%
%
The outer LDPC codes are constructed by lifting  the protographs using circulants that are optimized using the progressive edge-growth algorithm~\cite{PEG}.

\section{Numerical Results and Discussion}\label{sec:simresults}
In this section, we give AIRs and  frame error rate (FER) results for our designed optimized concatenated codes for the memory-$k$ nanopore channel model. 
We further compute AIRs and FER results for a real DNA channel using the experimental dataset in \cite{Srinivasavaradhan2021TrellisBMA}. The dataset  consists of $269709$ output  sequences taken from the output of an ONT MinION sequencer %
and also the corresponding input sequences (before synthesis).  There are in total $10000$ input sequences of length $N = 110$. 
In all our simulations, we use $L_{\text{max}} = 2$ and $d_{\max} = -d_\mathrm{min}  = 5 \sqrt{N \frac{\max\left(p_\I, p_\D\right)}{1-\max\left(p_\I, p_\D\right)}}$, where $p_\I$ and $p_\D$ are the average insertion and deletion probabilities based on the dataset in \cite{Srinivasavaradhan2021TrellisBMA}. Moreover, we use  separate decoding as described in Section~\ref{eq:separate}. The AIRs are computed by averaging over $10000$  sequences of length $N=110$.

\begin{figure}[t!]
 \centering \centering

 \includegraphics[scale = 0.7]{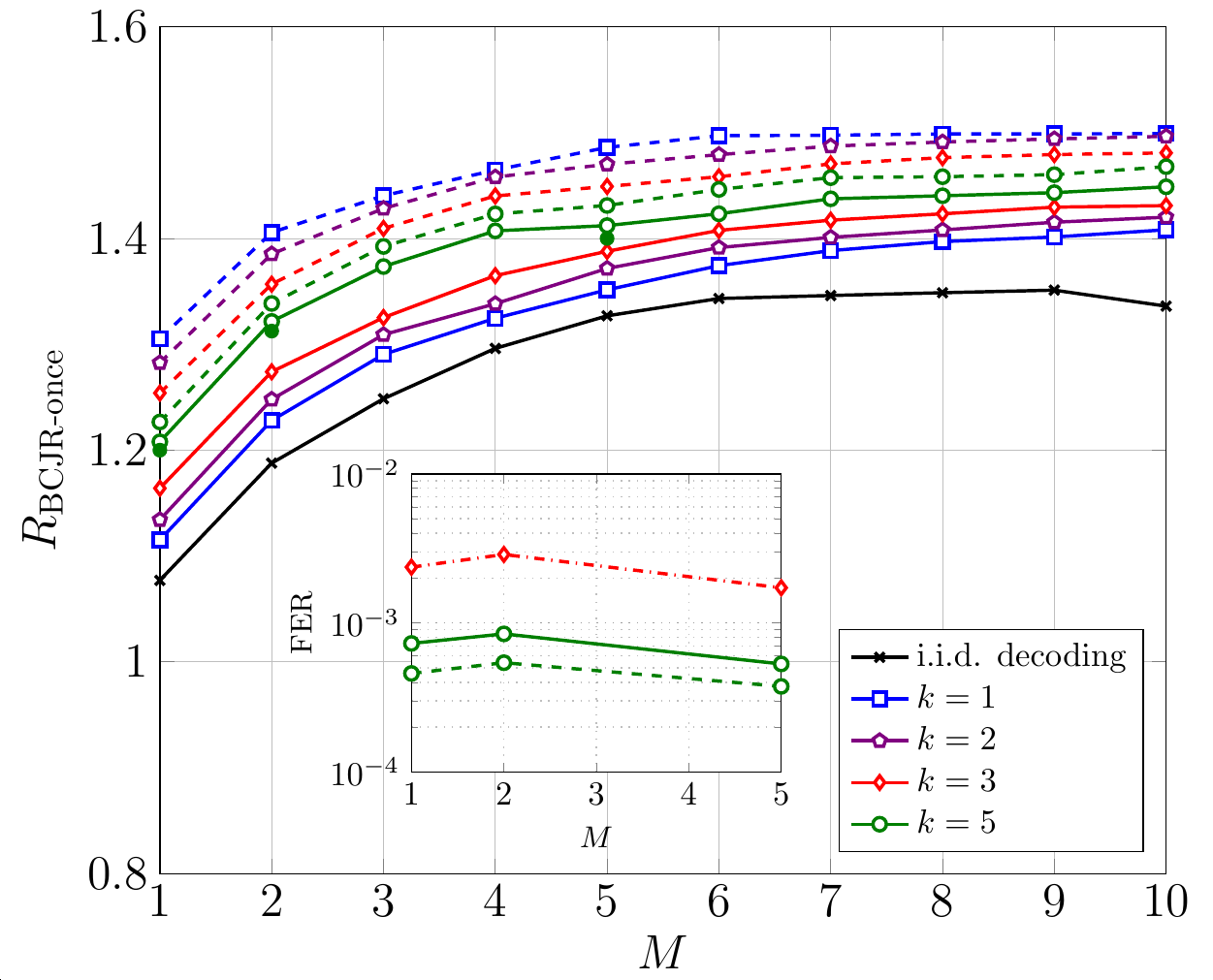}
	\vspace{-4ex}
	\caption{BCJR-once rates  with a rate-$\nicefrac{3}{2}$ inner convolutional code  for different values of $M$ and $N=110$. Solid curves are for the true channel and  dashed curves  for the memory-$k$ nanopore channel model. The dash-dotted line is for decoding with  $k = 3$ while the channel is set to $k = 5$. The green solid circles represent the rates for the outer LDPC codes with the optimized protographs in Table~\ref{tab:DE_protographs}.}
	\label{fig:BCJR-Once-Inner_DS_vs_Model_CC}
	\vspace{-2ex}
	 \end{figure}

In Fig.~\ref{fig:BCJR-Once-Inner_DS_vs_Model_CC}, we plot  BCJR-once rates for the  memory-$k$ nanopore  channel model  (see Section~\ref{sec:system}) with  the  convolutional code of Section~\ref{sec:inner_code} as a synchronization inner code (dashed curves) for different values of $k$. For each $k$, the inner decoder is matched to the combination of the inner code and the memory-$k$ nanopore channel model. %
Further,  for each $k$, we estimated the transition probabilities of the memory-$k$ nanopore channel model  as  described in Section~\ref{sec:model_build} using the dataset in \cite{Srinivasavaradhan2021TrellisBMA}.
We observe that the AIRs decrease with increasing $k$. %
This is expected, as increasing the memory $k$ makes the channel  more complex.

In the figure, we also plot AIRs for the true  DNA storage channel using the dataset in \cite{Srinivasavaradhan2021TrellisBMA} (solid curves). (The  random sequence added to the output of the inner code can be used to match the transmitted coded sequences to the input sequences of the dataset. Hence, the dataset can be used for the true channel with an inner convolutional code  as well.) In this case, the curve for a given value of $k$ corresponds to an AIR for the true DNA storage channel of a mismatched decoder where the inner decoder  is matched to the combination of the inner code and the memory-$k$ nanopore channel model. We observe the opposite effect to the AIRs for the memory-$k$ nanopore channel, i.e., the AIRs increase with increasing $k$. Again, this behavior is expected: If the memory-$k$ nanopore channel model models well the DNA storage channel (i.e., the assumption of a Markovian model is good), increasing $k$ makes the decoder better matched to  the true DNA storage channel, hence the AIR increases. Equivalently,  a low value of $k$ corresponds to a decoder that is more \emph{mismatched} with respect to the true DNA storage channel, resulting in a lower AIR. Our results hence support that this channel model is good. %
In fact, although not directly apparent from the figure, the AIRs saturate when $k$ increases, e.g.,  for the case of no inner code (results not shown here) the AIRs saturate for $k$ around $7$. Furthermore, interestingly, comparing the dashed and solid curves, we observe a sandwich effect, where the AIR curve for the memory-$k$ nanopore channel model (dashed curve with green circles) and the AIR for the true DNA storage channel with a mismatched decoder (matched to the combination of the inner code and the memory-$5$ nanopore channel model) are very  close. This indicates that increasing $k$ beyond $5$ does not bring much further gains in AIR for the true DNA storage channel.

In Fig.~\ref{fig:BCJR-Once-Inner_DS_vs_Model_CC}, we also plot the BCJR-once rate for the true DNA storage channel of a decoder that assumes i.i.d. IDS errors (black curve with star markers), as most decoders in the literature. We observe that this results in a significant performance loss.

Finally, in the figure, we also plot the rate $R$ obtained via DE for the true DNA storage channel for the outer LDPC codes with the optimized protographs in Table~\ref{tab:DE_protographs} (green filled circles),  showing that our  coding scheme gives  excellent performance at code rates close to the BCJR-once rates. %

In the inset of Fig.~\ref{fig:BCJR-Once-Inner_DS_vs_Model_CC}, we  plot the FER results of our designed optimized concatenated codes with the inner convolutional code of Section~\ref{sec:inner_code} and outer LDPC codes of length $\leno=10000$ based on the protographs in Table~\ref{tab:DE_protographs}. The outer codeword is split into $123$  subsequences of length $81$, resulting in $110$ channel input symbols after the inner encoding, and a single shorter subsequence of length $37$. %
The codes are simulated over both the true DNA storage channel with an inner decoder matched to the memory-$5$ nanopore channel model (solid curve)
and the memory-$5$ nanopore channel model for a decoder matched to $k=5$ (green dashed curve with circles) and $k=3$ (red dashed-dotted curve with diamonds). %
The results are in agreement with the DE results.%

The decoding complexity increases exponentially with  $k$ (see Section~\ref{sec:Complexity}). Thus, the AIRs and FER results in Fig.~\ref{fig:BCJR-Once-Inner_DS_vs_Model_CC} allow us to quantify the performance loss incurred by a decoder assuming a given memory $k$ and hence the trade-off between decoding complexity and error rate performance. We  observe that considering  memory $5$ incurs almost no loss in terms of AIR, indicating that $k=5$ is enough. 

\section{Conclusion}

We derived the optimum MAP decoder for  the memory-$k$ nanopore channel model. Based on the MAP decoder, we derived AIRs for the true DNA storage channel of a mismatched decoder that is matched to the memory-$k$ model and optimized coding schemes for this channel. We showed that, remarkably, the concatenated coding scheme in \cite{Maa22} (properly optimized) achieves excellent performance for the true DNA storage channel: Considering an optimal inner decoder for the memory-$k$ nanopore channel yields significantly higher AIRs\textemdash hence higher storage density\textemdash for the true DNA storage channel than a  decoder that assumes i.i.d. IDS errors, as usually assumed in the literature. 

%
%
%
%
%

%

%

\ifCLASSOPTIONcaptionsoff
  \newpage
\fi
\balance


%
%
%
%
%
%
%
%
%

\end{document}